\pdfoutput=1
\documentclass[10pt]{article}

\usepackage{color}
\usepackage{bbm}
\usepackage{epsfig}
\usepackage{amssymb,amsmath,amscd, mdwmath}
\usepackage{graphicx,cite,url}
\usepackage{algorithmic}
\usepackage{float}
\usepackage{amsfonts}
\usepackage{syntonly}
\usepackage{multirow}
\usepackage{graphicx}
\usepackage{subfig}
\usepackage{tabularx}
\usepackage{setspace}
\usepackage{stfloats}
\usepackage{amsthm}
\usepackage[labelfont=bf,labelsep=period,justification=raggedright]{caption}

\makeatletter
\renewcommand{\@biblabel}[1]{\quad#1.}
\makeatother

\date{}

\begin{document}

\begin{flushleft}
{\Large
\textbf{Modeling the coevolution between   citations  and coauthorship  of scientific papers}
}
\\
Zheng Xie$^{1, \sharp }$
Zonglin Xie$^{1}$
 Miao Li$^{2}$
Jianping Li$^{1}$
Dongyun Yi$^{1}$
\\
\bf{1}   College of Science,  National University of Defense Technology, Changsha, 410031,  China.
 \bf{2}~School of Foreign Languages, Shanghai Jiao Tong University, Shanghai, 200240,  China  \\  $^\sharp$ xiezheng81@nudt.edu.cn 
\\ The first three authors contributed equally to this work.
\end{flushleft}

\section*{Abstract}

Collaborations and citations within scientific research grow simultaneously  and interact dynamically.
Modelling the coevolution between them helps to study  many    phenomena
 that can be approached only through ¡¡ combining citation and
coauthorship data.
A  geometric graph for the coevolution is proposed, the   mechanism of which
  synthetically expresses  the interactive impacts of  authors and papers in a geometrical way.
The model is validated
  against    a  dataset of papers published on PNAS during 2007-2015.  The validation  shows the ability to   reproduce  a range of features observed with citation and coauthorship data combined and separately.
Particularly, in the empirical distribution of   citations per author  there exist  two limits, in which  the distribution appears as a generalized Poisson and a power-law respectively.
 Our model successfully reproduces the shape  of the distribution,  and provides an explanation for how the shape emerges via the decisions of authors. The model also captures   the empirically positive correlations  between the numbers of authors'   papers,
  citations  and    collaborators.


\section{Introduction}

Coauthorship    and citations  are typical  relationships in scientific activities, which can be  observed   from    datasets of scientific papers, and expressed as graphs
called coauthorship and citation networks respectively.
Coauthorship   networks, in essential, are    hypergraphs, in which
 nodes represent authors, and  the authors of a paper forms a hyperedge.
     Citation   networks are acyclic directed graphs, in which
papers are directly linked via   citation
relations.  Those networks provide   big pictures of the
topological connections between authors and papers respectively.
    Modelling the mechanisms
and temporal dynamics of those networks  (especially,   with   consideration of   the  papers' publication time) sheds light on   modes of collaborations within science,  and helps to discover  trends of academic research\cite{Glanzel1,Glanzel,Mali,Radicchi2}.




    Matthew effect describes the    phenomenon that success breeds success, which is
     also called preferential attachment or cumulative advantage\cite{Perc0}. The effect
        has been observed in  the    behaviors of   scientific collaborations and   citations   in specific disciplines and  countries\cite{Perc1,Kuhn,Perc2,Glanzel3}.
 As early
 as 1965, Price explained the emergence  of fat-tails  in the distributions of citation networks as a consequence of      Matthew effect.   His explanation is modelled by  a rule, namely  the probability of
  a paper's receiving a new citation is proportional to the citations it has received. The rule successfully predicts the scale-free property~\cite{Price,Price1}.
 Price's model has been generalized to illustrate other properties of citation networks  in various contexts~\cite{Radicchi1,Bornmann1,Evans,Goldberg,WangDS}. Meanwhile, although coauthorship networks receive  far less attention than citation networks\cite{Newman0}, there are still a range of   models
 working on the elementary
 mechanism for the emergences of
  scale-free, small-world  and degree-assortativity
  \cite{Barab,Moody,Perc,Wagner,Tomassini,Zhou,Milojevic,Xie3}.
    Each of those models has addressed Matthew effect.

The behaviors  of
 collaboration  and citation  are   coupled and coevolutionary.
Modelling the coevolution   has its own meaning, because  some    phenomena
  can be studied only by combining citation and
coauthorship data, such as   the     distribution of   citations per author,  self-/coauthor-/feedback-citation (a researcher  cites
others, who have  previously cited them)\cite{Martin ,Catanzaro}.

There has been relatively little work on modelling the coevolution of citation and coauthorship networks so as to simulate their simultaneous growth and dynamic interactions.  The first model (perhaps the only one according to our present knowledge)
is named TARL (topics, aging, and recursive linking), which successfully reproduces  a power law distribution of   citations per paper\cite{Borner1}.
The model novelly introduces  ``topics", which enables the simulated  citation networks to obtain positive and tunable  clustering coefficients.
The model     concentrates on the citation patterns, and   has validated the properties of the  citation network
  against   a dataset  of papers published on  the Proceedings of the National
Academy of Sciences (PNAS) 1982-2001.


To the ground, the behaviors  of
 collaborations  and citations are due to the self-organization  decisions made by authors: ``yes" or ``no". For example,  the
event that whether a paper cites another paper  can be treated as a
``yes/no" decision. Meanwhile, Matthew effect   ``is at the heart of self-organization across social and
natural sciences"\cite{Perc0}.
 A  geometric graph   is proposed
 to model those decisions and their Matthew effect.     The present model is built  on  a cluster of concentric circles, where each circle has a time stamp.
 The modelled decisions   synthetically  consider two  factors of generating collaborations and citations, namely   the  homophily (in sense of research interests, topics, etc.)
and the  academic impacts  (with  Matthew effect) of authors and papers. The model codes are available upon requests of readers.

 The model is  validated
  against   a   dataset  of papers published on
PNAS 2007-2015. The reasonability of the model is verified by the  respectable
model-data matching  in a range of  topological and statistical features of the empirical citation and coauthorship networks simultaneously, such as the distribution  of  collaborators/citations per author, and that of   citations per paper.   Those distributions  emerge  two limits, namely a generalized Poisson and a power-law in small and large variable regions respectively. There exists a cross-over  between the two limits.
Our model successfully reproduces the  shapes of those distributions,  and  reveals how the decisions of nodes in networks generate  the emerged limits.
 In addition,
the data PNAS  2007-2015 evidences the positive correlation between each  two of the three   author-indexes, namely  numbers of  papers,
    citations and  collaborators, which are also
captured
by our model.










This paper is organized as follows: the model and   data are described in Sections 2 and 3 respectively;  the      distributions of specific indexes per author, the correlation between each  two of the three   author-indexes  are analyzed in Sections 4 and 5 respectively; and the conclusion is drawn in Section 6.

\section{The model}

 The presented model
  adopts the viewpoint from research-teams in focusing on   the roles of research-teams performing in the production  and dissemination of knowledge. Each author is assigned   research-teams,   each paper is written by  a group of authors called   paper-team (which  is called   article team  in Reference\cite{Milojevic}), and the members of a paper-team   usually come from the same research-team.
 The model simultaneously generates two graphs interacting to each other, namely   a   hypergraph and a   directed acyclic graph,  to simulate the coevolution process of  coauthorship  and citation networks.

%

  The model is built on a cluster of concentric circles $S^1_t$, $t=1,...,T$.
For the sake of simplicity, the number of modelled  ``authors" is supposed to  linearly  grow  over  time  $t$. The parameter $t$  can be regarded as the $t$-th unit of time.
  Denote the $i-$th author-node   by $a(\theta_i,t_i)$, the $i-$th paper-node by $p(\theta_i,t_i)$, where $(\theta_i,t_i)$ is its spatio-temporal coordinate.

    Select  a small fraction of author-nodes  as ``leaders" of research-teams  to attach   zones to express their academic impacts~(called influential zones).
    For each
leader node, its research team is formed by the nodes within the influential zone of
the leader.
     All paper-nodes  are  attached  influential zones.
  The influential zones
are designed to  express the ability of capturing academic  communities'  response to authors and papers.
This design is built on the perception that counting  citations can offer   a  quantitative proxy of eye-catching ability.
 Note that counting  citations  is not a measure of the novelty  and  importance, which are   impossible to be measured objectively.
 The detail of   zones is given as follows.

Let constants  $\alpha_l>0$ and $\beta_l  \in[0.5,1]$ ($l=1,2,3$).
The zone of a leader  $a(\theta_i,t_i)$     is defined as an interval  of angular coordinate with   center $\theta_i$ and   arc-length ${\alpha_1} t^{-\beta_1}_it^{\beta_1-1}$.
The definitions of paper-node-zones are the same as those of leaders except  parameters   ($\alpha_2,\beta_2$  for  the papers written by   leaders as ``the first authors" and $\alpha_3,\beta_3$ for the papers written by non-leaders as ``the first authors").
 The zonal sizes   are all required to be  less than $2\pi$.
The reason of choosing  boundless circles is that  we do not need to consider spacial boundary effects on influential zones.

 The power-law factors    $t^{-\beta_l}_i$ ($l=1,2,3$) in influential zones induce  the scale-free property of synthetic networks~(see Appendix Eq.~\ref{eq3}).
The parameters of   zones  are used  to tune     synthetic  data  to fit the  empirical distributions of collaborators per author and those of citations per paper.
  The formula of zones is dependent on $t$. If  running the model with different initial time (which is equal to choose different $\alpha_l$),  the result will be different.



In the  empirical data,  the distributions of paper-team-sizes  and  references per paper    appear two common features, namely hook heads and
fat tails, which can be sufficiently fitted by generalized Poisson and power-law
distributions respectively~(which can be seen in following Fig.~\ref{fig11}a, Fig.~\ref{fig2}i).
For such kind of distributions, we  denote their probability density function  (PDF)
  by $f(x)$, $x \in  \mathrm{Z}^+$.

 After above preparations, we introduce the model   as follows, where the constants $N_1$,  $N_2$,  $N_3$, $T \in \mathrm{Z}^+$,   $p\in [0,1]$, and the paper-team-sizes and reference-lengths of paper-nodes are drawn from specific PDFs with form $f(x)$.
\begin{description}
\item[1.] Generate a coauthorship network
\item For time $t=1,2,...,T$ do:
\subitem  1.a.
  Sprinkle $N_1$  author-nodes as potential authors   uniformly and randomly   on   a   circle $S^1_t$, and  select $N_2$ nodes randomly from the new  nodes as leaders.
\subitem  1.b. For each new node $i$, search the   existing leaders whose zones cover  $i$, and join those leaders' research team.
For each   such leader $j$,  under probability $p$ ($p=1$ if  $i$   also is a leader),  generate a paper-team  with size $m$ by grouping together $i$, $j$  and $m-2$  research-team-members  of $j$  nearest to $i$, where   $m$   is   a random variable of a given $f(x)$  or      the research team size of $j$ plus one if the former is larger than the latter.
\subitem  1.c.  Select $N_3$ nodes  with non-zero degree randomly from all   existing  nodes, and generate a paper-team with size $m$   for each selected node $l$ by
grouping together $l$ and $m-1$ randomly selected  nodes with the same degree of $l$, where   $m$  is a random variable of  a given $f(x)$  or the number of nodes with the   degree  of $l$ if the former is larger than the latter.
\end{description}

\begin{description}
\item[2.]Generate a citation network
\item For time $t=1,2,...,T$ do:
\subitem  2.a.
   Each paper-team publishes a paper,   where ``the first author" is the new author-node   in Step~1.b  or the author-node
  selected  in Step~1.c.
 The spatio-temporal  coordinates  of   paper-nodes   are    those of their first authors.
 \subitem  2.b.
      Each  new paper      cites   the existing papers that have zones covering it, and then    cites  other existing papers uniformly at random (with no replacement) until the  length  of its references (out-degrees) equals   a random variable of  a given $f(x)$.
\end{description}
Note that in the publication  version    of this work (Scientometrics, 2017,  112: 483-507), there is expression error in  Step 1.b, where  ``neighbor" should be  replaced as ``research-team-member".

\begin{figure}
 \includegraphics[height=1.65 in,width=6.2   in,angle=0]{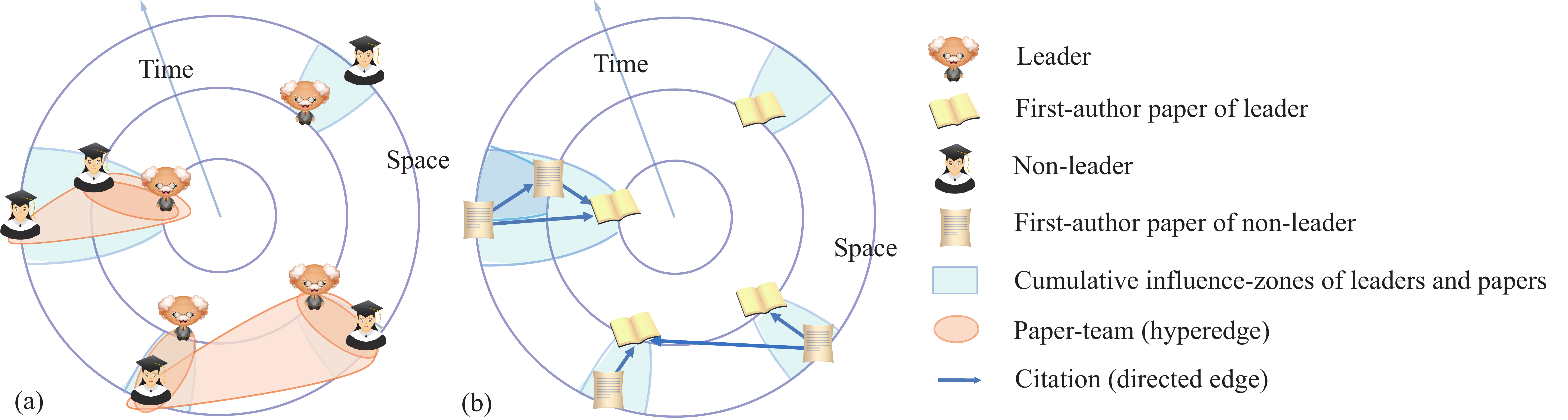}
 \caption{  {\bf Illustration of the model.} Leaders and papers have  influential zones, the sizes of which are all required to be  less than $2\pi$.   The rules of modelled authors publishing and citing  papers are outlined in the model. Those behaviors happen with in zones locally  in most cases, and across zones in few cases.
   The coordinates of papers are the same as those of their first authors.
} \label{fig0}      
\end{figure}



The homophily of citations and coauthorship  based on topics and research interests is a reason for the clear community structures in empirical data. In fact, with the development of sciences,    researchers   cannot understand all of the sciences, but focus  on their special fields,    collaborate with some of the researchers in the same   research-team, and cite some of the papers in their fields. This part of connections  is imitated by Step~1.b   and the first half of  Step~2.b, which contributes to the main part  of connections  in our model.

 The parameters of $f(k)$  in   Steps 1.b, 1.c and 2.b are  not required to be the same.
 The $f(k)$ in Steps  1.b and 1.c tunes  the distribution of paper-team sizes. The $f(k)$ in Step 2.b  tunes  the distribution of references per paper.
  Parameters $\alpha_1$ and $\beta_1$   tune   the distribution of collaborators  per author.
 Parameters $\alpha_{l}$ and $\beta_{l}$ (${l}=2,3$)  tune     the distribution of citations per paper.
There is  no     parameter  directly tunes the    distribution  of citations   per author.
 By far, we tune those parameters manually to fit empirical data. Firstly, we tune the parameters in Step~1 to fit the features of coauthorship data, then tune the parameters in Step~2 to fit the features of citation data. Lastly, we fine tune  the chosen parameters with an optimal trade-off between model-date  fits for the    features  of citation and coauthorship
data separately and
combined (e.~g. distribution  of citations   per author, self-citation rate). In the future, we will try to find   an effective way   to automatically learn    parameters from   empirical data.

If the authors only collaborate with the authors with the same research interests, and papers only cite the papers with the same topics, the networks of citations and coauthorship are highly clustered, but would not have giant components and the small-world property. However, it  is against   the empirical data, a   possible reason of which is the existence of    a small fraction  of  cross-disciplinary studies~\cite{Xie4}. Authors collaborating  across research-teams   would  result in
 the  connections of papers   from different topics  even disciplines
via citations.
  Step~1.c and the second half of Step~2.b  are designed to imitate the cross-disciplines phenomenon, which make  the modeled networks have the small-world property and   giant components. In addition,   researchers  can belong to different research-teams at different time, which  is also equivalently imitated   by Step~1.c   to some extents.


The model is partly based on our previous results.
The first part  of the model  is a generalization of    the coauthorship model in Reference~\cite{Xie6}.
The  parameter $p$ in  Step~1.b  is newly introduced   to tune   the average  number  of papers per author.
The second part of the   model comes  from the  citation  models  in References~\cite{Xie,Xie5}, but with  modifications, such as the increment of nodes, the connection rules, etc.

The innovation of the model here is  mapping the coordinates of     paper-nodes to  those of their first authors,  which connects papers to   authors and makes the networks of them  grow simultaneously and interactively~(Fig.~\ref{fig0}).
 The combination makes the new provided  model not only  reproduce  a range of statistical features of empirical coauthorship and citation networks (which have captured by our previous models\cite{Xie6,Xie5}),  but also  predict  some features   generated by the interactions between coauthorship  and citations, e.g. the distribution of   citations per author,
self-citation, etc.


%

There are some   oversimple assumptions in our model. (a) The   linear growth  of authors  does   not hold in reality. If changing it, the formula of zone sizes should be changed to  capture features of   empirical data. (b)  To express   degree assortativity (authors  prefer to collaborate with other authors with similar degrees),   nodes with the same degree  (the number of collaborators)  are grouped in    Step~1.c. Actually, no author does that surely when choosing coauthors. (c) The  authorship order in Step  2.a    does not consider heterogeneous conventions on different scientific  fields.

\section{The data}

 The information of citations  and collaborations  in  36,732   papers published  on PNAS during 2007--2015 is utilized to validate the model  in terms of major statistical  properties.  The empirical data are gathered from   the Web of Science (http://www.webofscience.com). Authors are identified by   their names on their papers. For example, the author named  ``Michael R. Strand" on his paper is represented  by the name per se.
Another way of  identifying authors    is  using   surnames and all   initials of authors~(e.~g. represent  ``Michael R. Strand"  by ``Strand, MR").
   Compared with using surnames and all   initials,
 using names on papers is a  reliable way to distinguish  authors
from one another in most cases. However, it   mistakes  one author as two if the author   changes  his/her name in different papers, and two authors as one if they have the same name.
Those deficiencies will cause the inaccurate of   expressing   real situations, such as  the number of    authors, the number of components,  clustering
coefficient, etc.


Several  references    say that the inaccurate (caused by using surnames and all   initials) does not much affect certain
research findings, such as
  the   distribution  type of  collaborators per author, and that type of  papers per author~\cite{Barabasi1, Milojevic1, Liben, Newman0,Martin}. Reference \cite{Kim1} analyzes   the  papers  in PNAS 2012, and shows   the   difference between the distribution  of  collaborators per the author identified by using surnames together with all   initials  and that by a  proxy of ground-truth. Those distributions have the same type approximately.
    Here we mainly focus on those distributions per se and some properties based on them. We  use  names on papers  to identify authors, because it seems more reliable~(Table~\ref{tab01}).   In elsewhere,  we will show the same
     mathematical laws (i.e.    types) underlying
    the considered distributions for the authors    identified by both methods  respectively, and analyze why this happens.
    \begin{table*}[!ht] \centering \caption{{\bf  Top ten authors (identified by using surnames together with all   initials  and  by using  names on papers respectively)   according to the number of papers/collaborators.} }
\footnotesize\begin{tabular}{l l r r rr r r r r r r r} \hline
 Papers   & Zhang  Y 137,
Wang  J  122,
Wang  Y  117,
 Liu  Y  112,
 Li  Y  104,
 Liu  J  95,
 Zhang  J  90,\\&
 Kim  J  89,
 Chen  J  80,
 Li  J  80 \\\hline
   Collaborators   &
 Wang  Y  1313,
 Zhang  Y  1297,
 Wang  J  1165,
 Liu  Y  984,
 Liu  J  856,
 Li  Y  816,\\&
 Zhang  J  762,
 Chen  Y  720,
 Li  J  695,
 Kim  J  688.
  \\\hline
 Papers  &
Nair  Prashant 62,
Croce  Carlo M 61,
Wang  Wei  52,
Mak  Tak W 49,
Onuchic  Jose N  45,\\&
Schultz  Peter G  44,
Ayala  Francisco J  41,
Langer  Robert  41,
Weissman  Irving L 41,\\&
Flavell  Richard A  40  \\\hline
Collaborators &  Wang  Jun  527,
 Croce  Carlo M  482,
 Wang  Wei  383,
 Langer  Robert  355,
 Schultz  Peter G  352,\\&
 Henrissat  Bernard  350,
 Wang  Tao  331,
 Wang  Ying  329,
 Mak  Tak W  328,
 Zhang  Yan  323    \\
\hline
 \end{tabular}
  \begin{flushleft}Those data come from    PNAS 2007-2015.
\end{flushleft}
\label{tab01}
\end{table*}

\begin{figure}
\includegraphics[height=2.3  in,width=6.   in,angle=0]{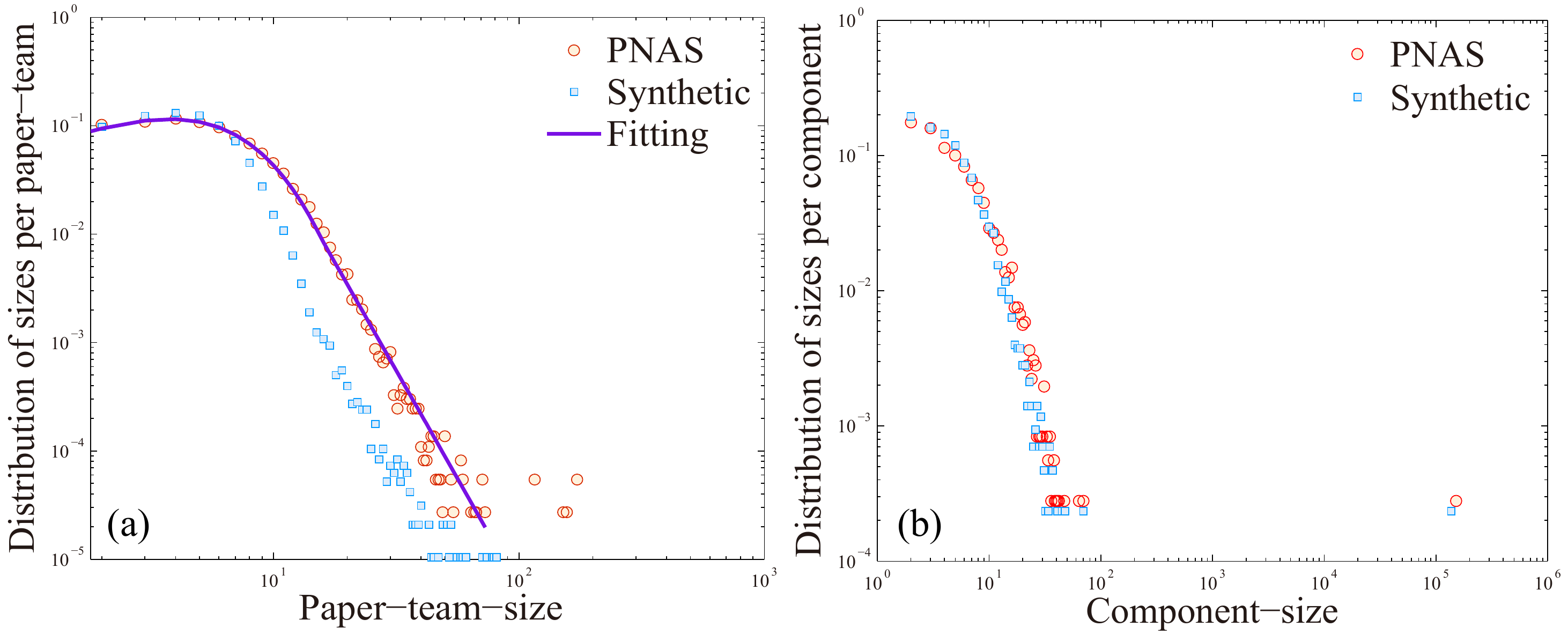}
 \caption{The distributions   of sizes per paper-team, and those per component.  Panels~(a, b) show  those distributions for
the empirical data PNAS 2007-2015 (red circles) and the   synthetic  data generated by our model with the  parameters in Table~\ref{tab0} (blue squares). The fitting  is a mixture of  generalized Poisson
and power-law distributions, the parameters of which are listed in Table~\ref{tab3}. }
 \label{fig11}      
\end{figure}

 Synthetic data are generated by the model to compare  with the empirical  data.
  The  model  parameters are listed in Table~\ref{tab0}.
   Denote the  PDFs of    generalized Poisson  and    power-law by   $f_1(x)= {a (a+bx)^{x-1}}{ \mathrm{e}^{-a-bx }/{ x!}} $ and $f_2(x)=cx^{-d}$   respectively,  where $a,b, c,d\in  \mathrm{R}^+$ and $x \in  \mathrm{Z}^+$.
Generate random variables of a distribution $f(x)$ with  head $f_1(x)$ and  tail $f_2(x)$ by
  sampling  random variables of $f_1(x)$ and $f_2(x)$   with probability  $q$ and $1-q$ respectively. In order to make $f(x)$ smooth  and    capture  the  empirical features at certain levels~(Fig.~\ref{fig11}a), we have tried   many times to  find   proper   $q$ and $f_i(x)$'s domain   $I_i$, $i=1,2$ for each step.


In PNAS  2007-2015, the maximum    collaborators per author and maximum citations per paper
are not very large.
So the values of $\alpha_{l}N_1$ (${l}=1,2,3$) should be also not very large~(Table~\ref{tab0}), which leads the probability of zone-overlapping  is  small.
The   paper-teams within and between
research-teams are modelled by Steps~2.b  and  2.c respectively. In reality, the  number of   paper-teams within
a  research-team is far more than   that between
research-teams.  So   $N_1$ is set to be  far larger than $N_3$.

\begin{table*}[!ht] \centering \caption{{\bf The parameters of the synthetic data.} }
\footnotesize\begin{tabular}{l llllllll  } \hline
Network sizes   & $T=4,500$ & $N_1=100$&$N_2=20$& $N_3=1$& $p=0.13$ \\
Influential zones&    $\alpha_1=0.133$& $\alpha_2=0.082$& $\alpha_3=0.02$& $\beta_1=0.42$& $\beta_2=0.44$& $\beta_3=0.44$    \\
$f(k)$ in Step  1.b &   $a=6$& $b=0$& $c=15,507$&  $d={-3.70}$& $q=0.968$&  $I_1=[2,250]$& $I_2=[11,150]$  \\
$f(k)$ in Step 1.c &  $a=6$& $b=0$& &   & $q=1$&   $I_1=[2,250]$&  \\
$f(k)$ in  Step  2.b &   $a=1.35$& $b=0.45$& $c=19,424$&  $d={-3.95}$ & $q=0.9999$&  $I_1=[0, 65]$&   $I_2=[20, 65]$   \\
\hline
 \end{tabular}
  \begin{flushleft}
\end{flushleft}
\label{tab0}
\end{table*}

In order to make   the synthetic data  capture  the  empirical indicators listed in Table~\ref{tab1} at certain levels,  we have attempted   many times to  find above parameters.
  Since the model is stochastic, we run 20 times with the same parameters, and   find that  the model is robust on those  indicators (Table~\ref{tab99}).

%
%
%


\begin{table*}[!ht] \centering \caption{{\bf Typical statistic   indicators of  the analyzed data.} }
\footnotesize\begin{tabular}{l r r r rr r r r r r r r} \hline
Network&NN&NE    & GCC &AP & MO & PG   &NG & AC  &SC & SC2\\ \hline
  PNAS-Citation &36,732&42,932&0.263 & 12.586 &0.922&  0.651 &9,145   &0.287 & 0.312&0.399\\
  Synthetic-Citation &95,993&133,318 & 0.136  & 8.034 &0.713 & 0.807 &16,128     &0.280 & 0.231 &0.388\\
  PNAS-Coauthorship &162,531 &1,074,836 & 0.896 &6.662 &0.909&  0.844 &5,013  & 0.554 \\
  Synthetic-Coauthorship &175,699 &886,140 &0.825 & 6.808 &0.948 & 0.864 &3,578 &0.164    \\
\hline
 \end{tabular}
  \begin{flushleft} The indicators are  the numbers of nodes (NN) and edges (NE),  global  clustering coefficient (GCC),  average shortest path length (AP),  modularity (MO),  the node proportion of the giant component~(PG), the number of components (NG), the assortativity coefficient of degrees for coauthorship networks  and that of in-degrees for citation networks~(AC), the proportion of self-citations (SC) and that of citations by collaborators (SC2).
  The  AP of   the last two networks is calculated by sampling $15,000$  pairs of nodes respectively.
\end{flushleft}
\label{tab1}
\end{table*}

\begin{table*}[!ht] \centering \caption{{\bf Specific   indicators' mean  and   standard deviation (SD).} }
\scriptsize \begin{tabular}{l r r r rr r r r r r r r} \hline
Synthetic&NN&NE    & AC &GCC & PG & NG   &MO & AP  &SC & SC2\\ \hline
 Citation\\
Mean& 9.6E$+04$&1.3E$+05$&2.7E$-01$&1.3E$-01$&8.1E$-01$&6.8E$+00$&7.1E$-01$  &8.2E$+00$	  &2.3E$-01$&3.9E$-01$\\
 SD &
3.4E$+02$&
5.3E$+02$&
8.9E$-03$&
1.1E$-03$&
1.8E$-03$&
1.8E$+02$&
1.4E$-03$&
 1.2E$-01$&
1.2E$-03$&
2.7E$-03$\\
 Coauthorship\\
Mean&
1.8E$+05$&8.7E$+05$&1.7E$-01$&8.2E$-01$&8.6E$-01$&3.5E$+03$&
9.5E$-01$&6.8E$+00$\\
 SD &8.0E$+02$&8.7E$+03$&2.7E$-02$&8.1E$-04$&2.7E$-03$&4.5E$+01$&
6.5E$-04$&2.1E$-01$\\
\hline
 \end{tabular}
\footnotesize
   \begin{flushleft} The meanings of   headers are shown in the notes of Table~\ref{tab1}.
\end{flushleft}
\label{tab99}
\end{table*}


Besides the statistical features in next sections,
our model   captures three    topological features of the empirical data (Table~\ref{tab1}), namely giant component (Fig.~\ref{fig11}b),  clear community structure (high   modularity) and small world (average shortest path length~$\sim\log$(the number of nodes), positive   global clustering coefficient). In the model,
the nodes in the same research-team probably belong to the same  community.  Setting $N_1\gg N_3$ makes   edges within research-teams are significantly more than those between research-teams, and consequently   makes  the synthetic networks have a clear community structure.
The positive     global clustering coefficient  due to the homophily in the sense of geometric distances. The small  average shortest path length  is caused by the random connections generated in Step 1.c and in the second  half of Step 2.b.

\begin{figure}
\includegraphics[height=2.3  in,width=6.   in,angle=0]{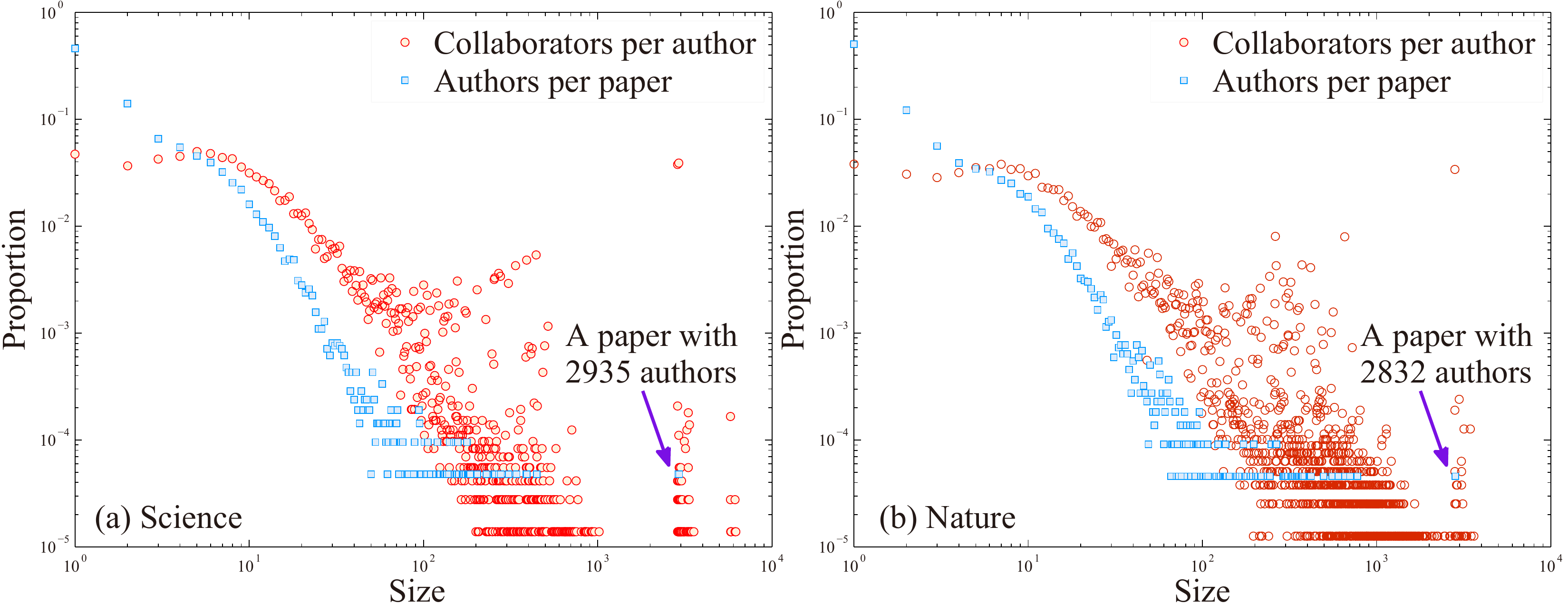}
 \caption{{\bf The distributions   of collaborators per author and those of authors per paper-team.} The data in Panels (a, b) are extracted from the papers published on Science and Nature during
2007-2015 respectively, which are gathered from   Web of Science.  The authors are identified by their names on papers.   }
 \label{fig12}      
\end{figure}

More than 80\%  papers of PNAS 2007-2015  belong to biological sciences.
In Appendix, in order to show    the  model's flexibility,   we fit    another dataset in physical sciences.
    The model is unsuitable for the datasets having  very large paper-team-sizes, e.~g.
     the   papers of Nature and Science published during 2002-2015.  The papers with very many authors    cause  that many authors have many collaborators, and consequently  the distributions   of collaborators per author have no power-law tail~(Fig.~\ref{fig12}).
\section{Modeling the distributions of specific  indexes per author}

\begin{figure}
\centering
\includegraphics[height=6.6    in,width=6.2   in,angle=0]{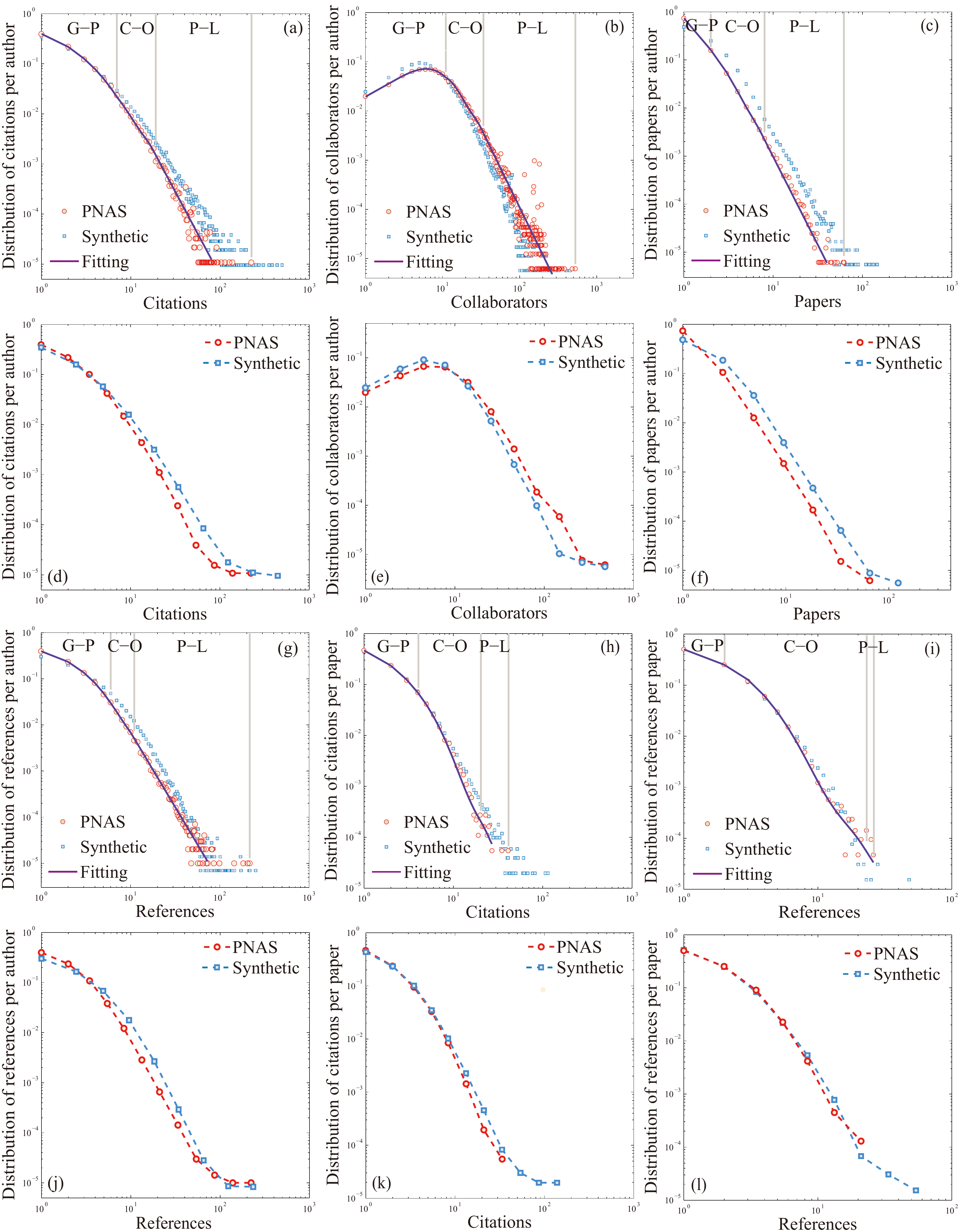}
 \caption{ {\bf The empirical  and synthetic   distributions of  collaborators/citation/papers/ references per author,  and those of  citations/references per paper.} The citations coming from or citing  the papers (which are not contained in the empirical data)  are not   counted. In Panels (a-c, g-i), the regions ``G-P",  ``C-O", ``P-L" stand  for generalized Poisson, cross-over  and power-law respectively. The parameters and goodness of fittings are listed in Table~\ref{tab3}.
    In Panels (d-f, j-l), the data are binned
on abscissa axes  to     extract the trends
   hiding in   noise tails. }
 \label{fig2}      
\end{figure}

The distribution   of  collaborators per author $P_{AA}(k)$,   and the distribution  of citations  per author $P^-_{PA}(k)$/ per paper    $P^-_{PP}(k)$  ($k>0$)  share a common characteristic: a generalized Poisson head, a power-law tail and  a cross-over between them~(the purple lines in Fig.~\ref{fig2}). Note that the cases of $k=0$   are not considered, because   solitary nodes   do not affect   network topology.
 The presented geometric   model provides a  respectable model-data fit~(the  blue  squares  in Fig.~\ref{fig2}).
    With some modifications,  the  analysis and calculations  in References~\cite{Xie,Xie1,Xie3,Xie5,Xie6}  can be employed to show how the model works, which are    described in Appendix. Here
  an intuitionistic   explanation is given as follows.

\begin{table*}[!ht] \centering \caption{{\bf  The parameters and goodness   of fitting functions.} }
\begin{tabular}{l r r r rr r r r r r r r} \hline
Distribution   &$a$ &$ b$   & $c$ &$d$  &$s$ & $B$   & $E$   & $G$ &$P$  &$p$-value & Sum \\ \hline
Citations per author & 0.810  &0.465 & 27.88&3.337  &   1.747    & 7& 19 & 6& 30& 0.548&0.998  \\
References per author&  0.960& 0.384  &12.29 &3.245 &1.584 &6 &11    &10&15&  0.395 & 1.003\\
Collaborators per author &   4.916&  0.463 & 294.1 &3.213 &  0.858 &  11 &34 &12&45 &    0.082
& 1.004 \\
 Papers per author & 0.703 &0.387&21.79 &3.821&1.968 & 2 &8 &10&25& 0.038  & 0.995\\
  Citations per paper  &0.703 &0.387 &31.89 &3.942& 1.968  & 4 &20&10 &20&       0.994

& 0.999\\
References per paper  &  0.768& 0.276 &23.48&4.131&    1.856& 2 &23 &12&15&0.873&   1.005\\
Sizes per paper-team  &3.657&0.402  &532.4 &3.989&   1.194& 15 &16 &10&20&0.188 &1.000 \\
\hline
 \end{tabular}
  \begin{flushleft}
  The ranges of  generalized Poisson $f_1(x)$,  cross-over, and power-law $f_2(x)$  are $[1,E]$,  $[B, E]$, and  $[B,\max(x)]$ respectively.
  The fitting function is $f(x)=q(x)      s f_1(x) +(1-q(x))f_2(x)$, where
  $ q(x)=\mathrm{e}^{ - (x -B)/(E-x ) }$.    The fitting processes are:
  Obverse proper $G$ and $P$; Calculate parameters of $sf_1(x)$ (i.e. $a$,  $b$,    $s$) and $f_2(x)$ (i.e. $c$,    $d$)   through regressing  the empirical  distribution   in  $[1,G]$ and $[P,\max(x)]$ respectively;      Find  $B$ and $E$ through exhaustion    to  make $f(x)$ pass  Kolmogorov-Smirnov  test ($p$-value$>0.01$).  The $p$-value   measures the  goodness-of-fit.
 The sum  of $f(x)$ over  $[1,\max(x)]$    near-unity shows that $f(x)$ can be regarded as a probability density function.
   \end{flushleft}
\label{tab3}
\end{table*}

Behaviors of collaborations  and citations   are    dependent on the choices of authors, the attractiveness of authors and papers.   The choices can be simplified by  ``yes/no" decisions.   Take citation behavior as an example.
Treat  the event that whether a paper cites another paper   as a  ``yes/no"    decision. Then  the  number of a paper's citations     is the number of successes in a sequence of $n$    decisions, where $n$ is the number of the papers having willing to cite that paper.
Approximate   the    probability $p$ of ``yes"  by its expected value $\hat{p}$, and suppose those ``yes/no"  experiments  are independent.
  Then,  the number   of a  paper's  citations will follow  a binomial distribution  $B(n,\hat{p})$.
When   $n$ is large and $\hat{p}$ is   small,
  $B(n, \hat{p})$ can be approximated by a Poisson distribution  with mean   $n\hat{p}$.
An author could publish several papers.   So the number of an author's citations   is the sum of several random variables drawn from Poisson, which is still  drawn from Poisson. Note that the result holds  under above simplifying assumptions.

  The ``yes/no"    experiments   could be affected by previous occurrences, e.~g. two papers written by the same authors highly probably share some references.
Meanwhile, the values of  $p$ and $n$ are    not    constant due to the  diversity of  attractive abilities of papers and authors. Hence   it is  reasonable    to think   the number of citations received by lowly cited authors and papers are drawn from
     generalized Poisson distributions,  which relax the restrictions  of    Poisson distribution by
  allowing   the probability of occurrence of a single event to affect  by previous occurrences\cite{Consul}.
     For   highly cited papers (with  large $np$) and authors,  the numbers of    citations  are   large enough to suppose the ``yes/no"    decisions are independent. So  the number of  citations could be considered as random variables  drawn from a   range of
      Poisson distributions with  sufficiently large expected values.

The  diversity of  attractive abilities of papers  gives the possibility of   existing   papers with  highly attractive abilities,
and then guarantees  the  relative  commonness for the existence of  papers getting citations that greatly exceed  the average. The commonness appears as  the emergence of   citation distributions' fat tails. The diversity of attractive abilities induces   various probabilities  of ``yes" $p$ and various numbers of potential citations $n$.  In the model, those are generated by the power-law factor   in the   formula   of influential zones' size,  namely $t^{-\beta_l}_i$, where $l=1,2,3$, and  $i$ refers to  specific node.
 The factor gives a sufficient diversity for the attractive abilities of nodes, and consequently derives the scale-free property of   modelled networks.
  Eq.~\ref{eq15} in Appendix shows how it works, which illustrates
     the derivation   of power-law by averaging  a range of  Poisson distributions with expected values proportional to  ${s}^{-\beta}$ ($s=1,...,T$, $T\in \mathbb{Z}^+$).
      Actually,
 in systems science,   diversity is  often regarded to be a reason for complexity, and scale-free is one of the symbols of complexity~\cite{Christensen}.




Note that there exists a difference between BA model\cite{Barab} and our model. In BA model,   nodes make decisions to link previous nodes based on understanding of all  nodes' degrees. In our model, most decisions made by  nodes are  local behaviors, which are restricted by geometric locations~(Fig.~\ref{fig0}). Meanwhile, a few decisions are made randomly, which are modelled by Step~1.c and the second half of Step~2.b. In reality, authors make decisions based on   their   knowledge. So the decisions are affected by the locality of authors' knowledge.   Uncertainty  is also in authors' decisions. In our model, the connection mechanism addresses   the locality as well as the uncertainty.

In statistics,  mixture distributions, e.~g. $P^-_{PA}(k)$, mean   samples come
from different populations.
   In reality,  the main part of  the  authors in scientific papers  is composed of the teachers and students in institutes or universities, which can be treated as two   populations.
 Research modes of students and  teachers are   different. Many students only write a few papers, and do not write after graduations, but their teachers could continuously write papers  collaborating with   new students or   researchers, and  so persistently receive citations.
Meanwhile,  due to the aging of papers, the citations received by students  cannot persistently increase  on average. 


In our model, the leaders are designed to play  the role as teachers, and other team members as students.
Note that the   distribution  $P^-_{PA}(k)$ of the empirical data  is   not perfectly fitted by that of the synthetic data.
However,  the  respectable model-data
fit  after data binning  is still
impressive  to us  because the model  involves no true free parameters to tune $P^-_{PA}(k)$. It also confirms the reasonability of the above analysis.
In addition, a similar  analysis   has been applied    to $P_{AA}(k)$~\cite{Xie6}.




The smoothness  of $P^-_{PP}(k)$ and $P^-_{PA}(k)$ does not appear in $P_{AA}(k)$. In reality, the papers of an author and the   citations of a paper   increase smoothly over time. However,   a paper with very many authors can  make those authors' collaborators   increase rapidly.




\section{Modeling the Matthew effect in academic societies}



\begin{figure}
\includegraphics[height=3.3  in,width=6.3   in,angle=0]{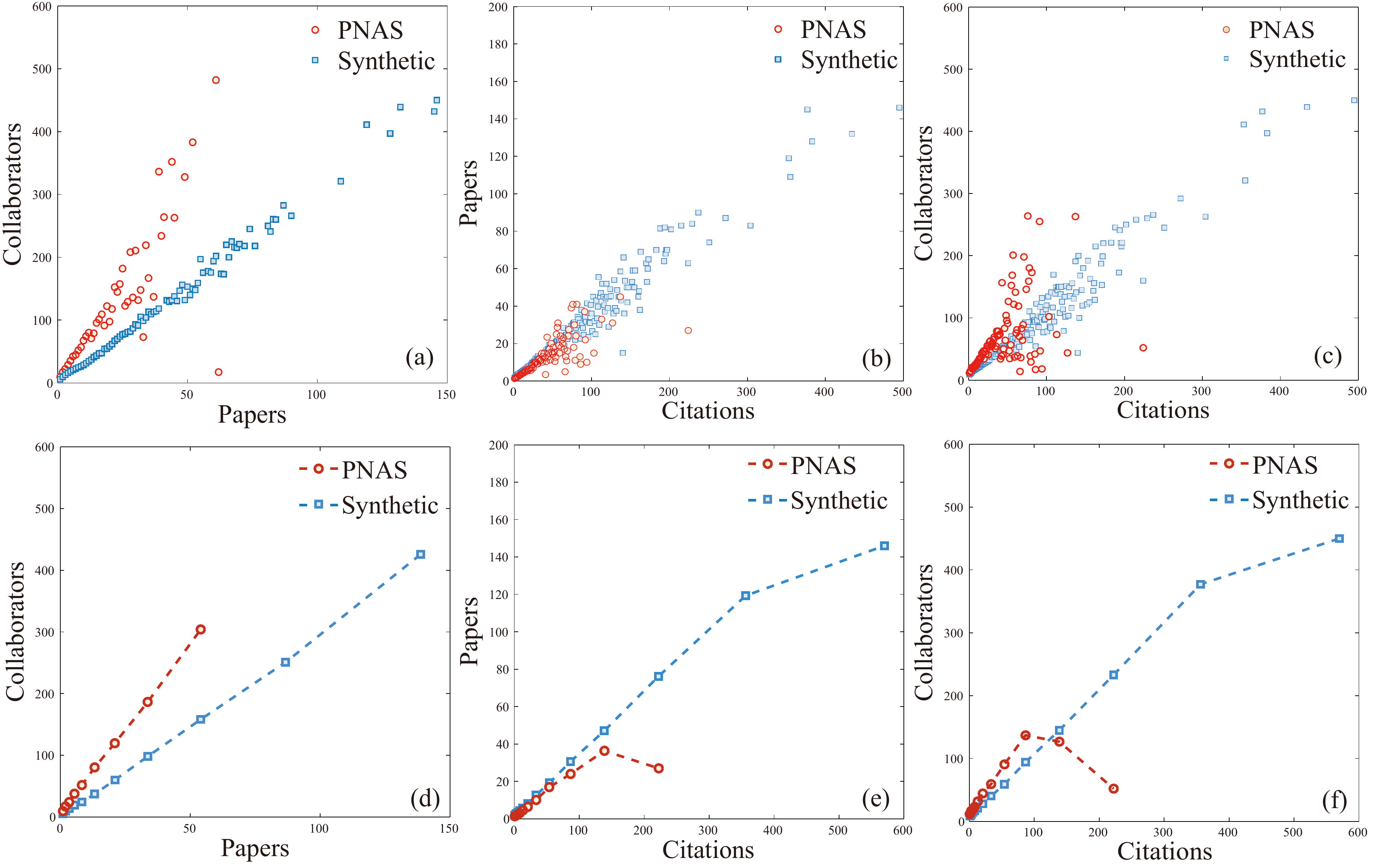}
 \caption{{ \bf  The positive correlations between three  author-indexes:   numbers of   papers, citations and collaborators. }  In Panels (a-c),  each point $(x,y)$ means  an author published $x$ papers   has $y$ collaborators  on average.
    In  Panels (d-f), the data are binned
on abscissa axes to   improve visibility for the positive slopes.}
 \label{fig4}      
\end{figure}

From the social viewpoint,    Matthew effect  (the rich get richer)  naturally exists
 in academic fields.  Authors with many citations and papers  can  improve their chances of attracting collaborators, especially outstanding students.
 Consequently, those
authors may write more
high-quality papers and increase their chances of receiving
citations. So  the indexes of authors, namely numbers of collaborators, citations and papers, improve  mutually  and form a positive feedback loop, which is evidenced partially by
the positive correlations between those indexes~(Fig.~\ref{fig4}). Funding plays a part as  an activator in promoting the   feedback loop and  Matthew effect\cite{Borner1}.  The authors with voluminous papers and   citations easily obtain funding, which  in return enables those authors to obtain  more  collaborators (especially postdocs and visiting scholars), citations,   papers
and hence   more funding.


Matthew effect is often regarded as an    explanation  for the scale-free properties of  many real networks.
Here we reverse the question:
Does there exist any  Matthew effect   in the developments of the authors with a kind of the aforementioned  indexes  following power-laws?
With the information of papers' publication time,  such  Matthew effect can be observed   directly as follows.

\begin{figure}
\centering
\includegraphics[height=3.3  in,width=6.2   in,angle=0]{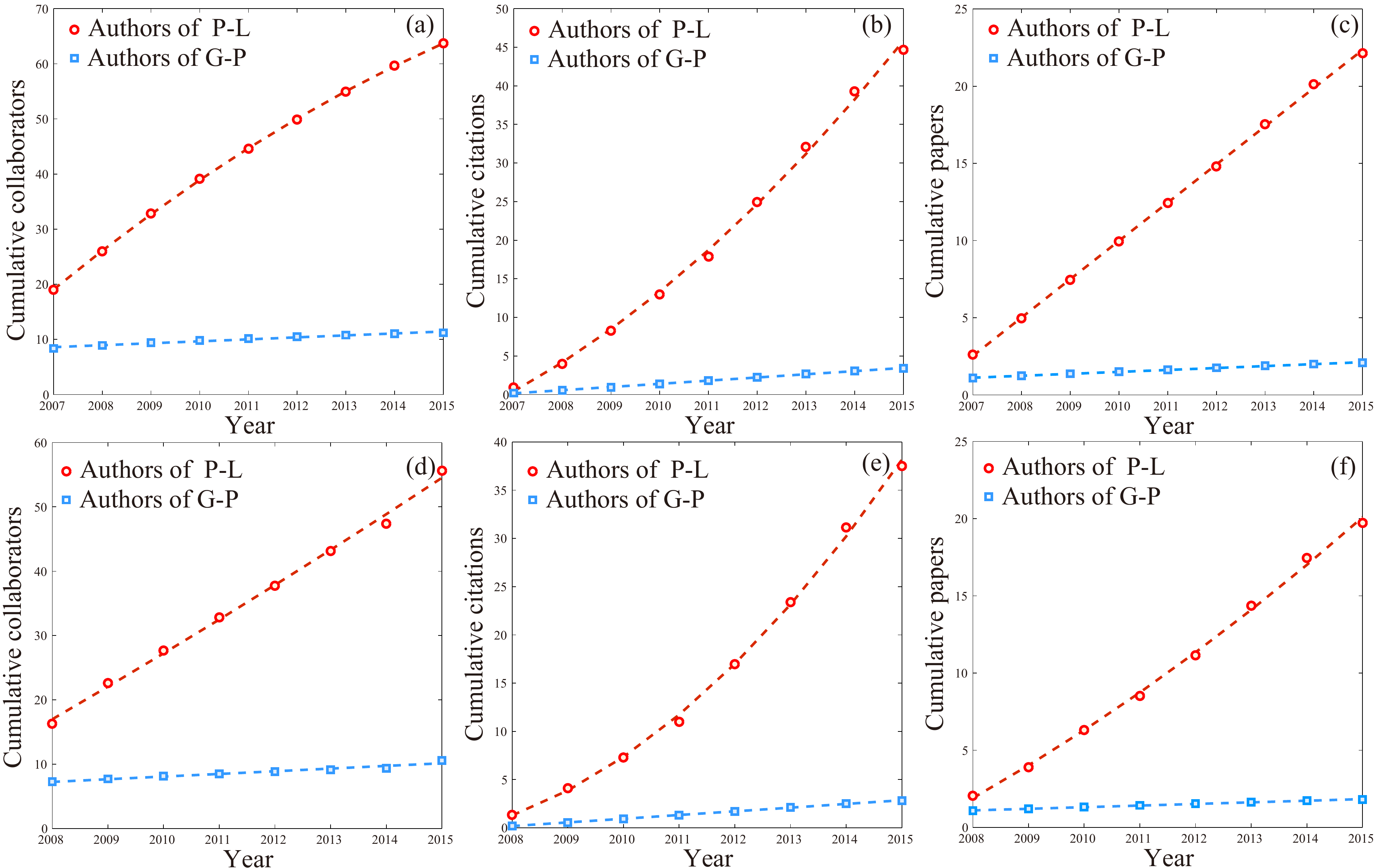}
 \caption{ {\bf  The  Matthew effects about   three author-indexes:   numbers of   papers, citations and collaborators.}
 Authors of G-P and P-L are the subsets of the authors, whose   indexes   follow  a generalized Poisson and power-law distribution respectively.
 The  points  (33/29/15) splitting the distributions of collaborators/citations/papers per author   into  G-P  and  P-L    are
  detected by the algorithm in Table~\ref{tab4}.
   The data in panels (a-c) and panels (e-f)  are about the authors first appearing in 2007 and 2008 respectively. The fitting curves are quadratics.
} \label{fig3}      
\end{figure}

For each kind of indexes, split   authors into two   non-overlapped parts by the algorithm  in Reference~\cite{Xie6}~(Table~\ref{tab4} in Appendix), namely a  generalized Poisson part and a power-law part. The indexes' cumulative values~(over time)   of the authors in the two parts  follow generalized Poisson and power-law distributions  respectively, which are proved by the Kolmogorov-Smirnov tests for the fittings in Fig.~\ref{fig2}~(a-c). The indexes of the authors in the  generalized Poisson part  grow quite slowly on average, meanwhile,
those in the power-law part grow fast, even in   accelerated ways for   citations~(Fig.~\ref{fig3}). It evidences that  the phenomena ``the rich  getting richer" has  existed in   academic
societies.



The Matthew effects about  the   aforementioned    indexes are coupled due to the positive feedback loop and   positive correlations between those indexes.
 Our model gives a geometrical  expression of   those coupling Matthew effects.  Older   leaders have a larger influential zone, so more easily capture collaborators, and then   publish more papers to receive more citations. Therefore, our model successfully   reproduces the power-law tails in the   distributions of those indexes~(Fig.~\ref{fig2}a-c), and   positive correlations between those indexes~(the positive slops in Fig.~\ref{fig4}).
In essentially,  the Matthew effect in our model  is a cumulative advantage on time.



\begin{figure}
\centering
\includegraphics[height=1.7  in,width=5.8  in,angle=0]{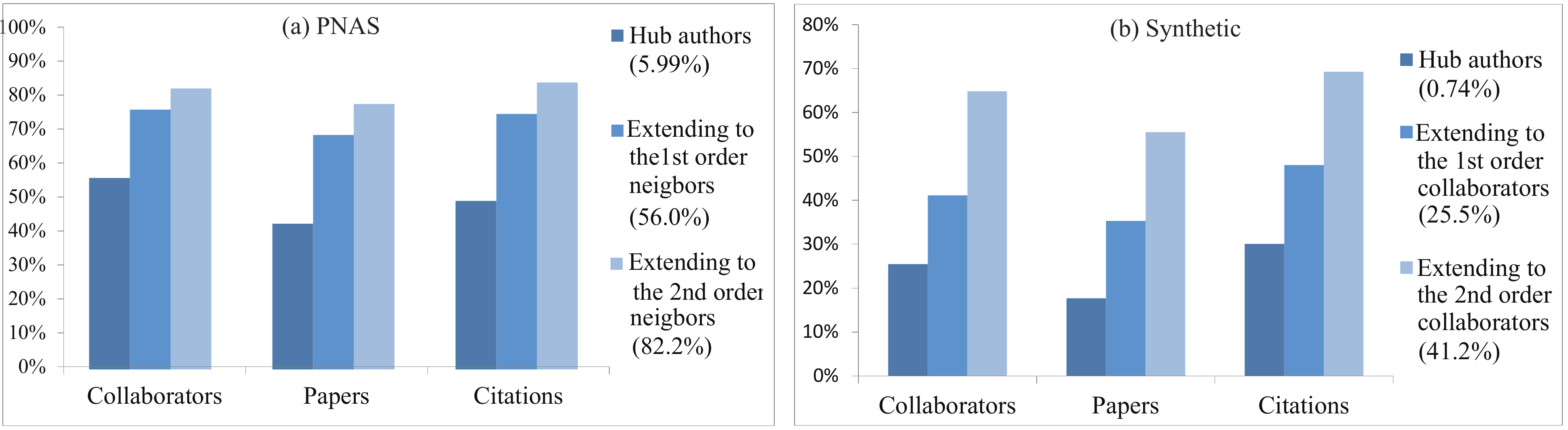}
 \caption{ {\bf   The proportions of papers, citations, collaborators obtained by  hub-authors and  their  collaborators.}
Hub-authors  are those   with collaborators more than the cut-off points   detected by the algorithm in Table~\ref{tab4} (55 for PNAS and 56 for the synthetic data).   }
 \label{fig9}      
\end{figure}

  Matthew
effect will lead to   ``the strongest  taking  over" in academic  societies, which  has  emerged in the empirical data~(Fig.~\ref{fig9}).
  There exist 5.99\% (9,742/162,532) authors (called hub-authors), whose numbers of collaborators (more than 32)   follow a power-law distribution, published $42.6\%$ papers and received $49.3\%$ citations.  The hub-authors and their   collaborators    obtain nearly 80\%   papers and citations. In addition,
the topological relationships  between   hub-authors  are very close. The proportion of a hub-author   having   coauthored with another hub-author  is $99.5\%$ ($9742/9697$).

For each hub-author,
 we calculate the  gap  between the   first  appearance  time of him/her  and  each of  his/her hub-author-collaborators, and record  the    maximum gap.
The mean of those maximum gaps
is $ 4.180$ years.
   Replace the condition that  the number  of collaborators is more than $32$ by that the number of citations is more than $28$ (the number of papers is more than $14$). Then the proportion
    of an author satisfying the condition having
coauthored with another  author also satisfying the condition
    is $352/401= 87.8\%$ ($243/285= 85.3\%$), and the mean of the maximum time-gaps   is $1.104$ ($ 1.203$) years.
Those evidence  that  Matthew effect has been inherited, and works  on research-teams not just on individuals.


From the viewpoint of systems science,   ``the strongest  taking  over"
will suppress    system diversity, and consequently harms    system flexibility.
Hence there exist two critical strategic problems for research administrators: How to reform  the existing  academic evaluation methods  and  funding  mechanisms,  which are mainly oriented by specific  indexes with Matthew effect, e.~g. the number  of citations; How to design a   reasonable inspiriting mechanism of scientific research to protect   academic diversity.

Some literatures discussed the reason and consequences of  Matthew effects in academic  society.
Reference\cite{Squazzoni } says
Matthew effects are expected if scientific community competes for the same attention device, e.~g.  a top journal. However, Matthew effects  could  probably erode the competition, so have negative implications on innovation\cite{Balietti}. For example,   the most competitive journals of  medical sciences rejected  some important articles. As Reference\cite{Balietti} says,
 ``tolerating early failure and focusing on long-term
success  could be a fine way to guarantee high level  innovations".

\section{Conclusion}

The presented model adopts a
viewpoint  in focusing on the coevolution of  collaborations and citations within scientific research, and particular  in
  reproducing statistical and  topological features generated  by
 the interactions between collaborations and citations.
The   good  model-data
fitting shows the reasonability of the model.
 Especially,
 the model    provides a respectable prediction of   the empirical distribution of
citations per author.   The model potentially paves a geometric way to   figure out
the interactional modes of collaborations and citations.
For example, it explains
 the emergence of  generalized Poisson and Power law in
the distribution of
citations per author    by the different collaboration modes of teachers and students.

      Some shortcomings of the model are indicative
of the need for further research: The increment of new nodes should not be  fixed;  More reasonable expressions are needed for the academic  communications between research-teams.
We are especially
interested in the self-similarity emerged in the empirical data, such as the distribution  of  citations per paper, and that of   collaborators per author.
  We also should   discuss
 the role of the coevolution  in the formation and evolution  of academic  communities.







\section*{Acknowledgments}
We thank Professor   K. Christensen    for the  valuable suggestions on the description of ``cross-over", Professor  J.~Y.~Su  for proofreading this paper.  This work is supported by the   fund
from  the national university of
defense technology teacher training  project
(No. 434513512G).

\section{Appendix}

\subsection{Detecting boundary   for probability density functions.}

The boundary    detection  algorithm for probability density functions (PDF) is listed in Table~\ref{tab4}, which   comes from  Reference~\cite{Xie6}.
\begin{table*}[!ht] \centering \caption{{\bf A boundary detection algorithm for PDF} }
\begin{tabular}{l r r r r r r r r r} \hline
Input: Observations  $D_s$, $s=1,...,n$,  rescaling function $g(\cdot)$, fitting model     $h(\cdot)$.\\
\hline
For   $k$ from $1$ to $\max(D_1,...,D_n)$ do: \\
~~~~Fit   $h(\cdot)$  to   the PDF $h_0(\cdot)$ of    $\{D_s, s=1,...,n|D_s \leq k\}$    by  maximum-likelihood\\ estimation; \\
~~~~Do   Kolmogorov-Smirnov (KS) test for two    data
     $g(h(t))$ and $g( h_0(t))$, $t=1,...,k$ \\
   with the null hypothesis  they coming from the same continuous distribution;\\
~~~~Break  if  the test rejects the null hypothesis  at     significance level $5\%$. \\ \hline
Output: The current $k$ as the   boundary point. \\ \hline
 \end{tabular}
    \begin{flushleft}
    \end{flushleft}
\label{tab4}
\end{table*}
\subsection{Simplifying the model}

\begin{figure}
\includegraphics[height=1.5  in,width=6  in,angle=0]{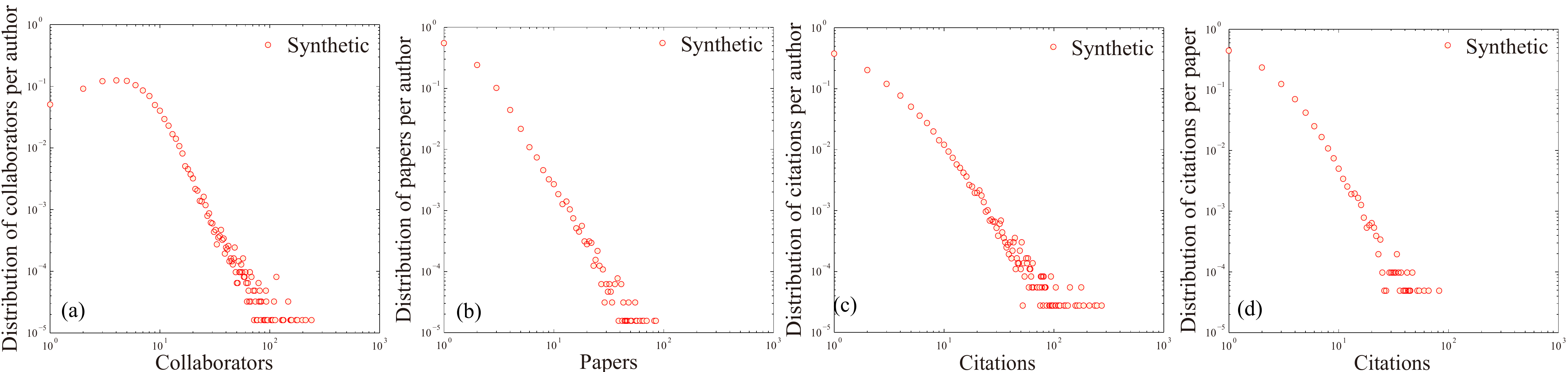}
\caption{  {\bf  The   synthetic   distribution  of  collaborators/citations/papers   per author,   and that  of citations   per paper.  } The model parameters are listed in Table~\ref{tab6}.
   } \label{fig12a}   
\end{figure}

An obvious    weakness of the provided model is that it has a lot of parameters.
 If ignoring the fitting of the   distribution  of references per paper and that of paper-team sizes,   we can reduce the model's parameters as those in Table~\ref{tab6}. The reduction   does not affect the synthetic
  distribution  type of collaborators/papers per author,  and that  of citations per paper/per author~(Fig.~\ref{fig12a}).
\begin{table*}[!ht] \centering \caption{{\bf The parameters of the synthetic data.} }
\normalsize\begin{tabular}{l llllllll  } \hline
Network sizes   & $T=4,500$ & $N_1=100$&$N_2=20$& $N_3=1$& $p=0.13$ \\
Influential zones&    $\alpha_1=0.133$& $\alpha_2=0.082$& $\alpha_3=0.02$& $\beta_1=0.42$& $\beta_2=0.44$& $\beta_3=0.44$    \\
$f(k)$ in Step  1.b &   $a=5$& $b=0$  \\
$f(k)$ in Step 1.c &  $a=5$& $b=0$&  \\
$f(k)$ in  Step  2.b &   $a=1.35$& $b=0.45$&    \\
\hline
 \end{tabular}
  \begin{flushleft}
\end{flushleft}
\label{tab6}
\end{table*}

\subsection{The underlying formula for the   distribution of citations per paper}

We only analyze the underlying formula for the distribution type of synthetic  ``citations per paper" (in-degrees), which is similar to that in References~\cite{Xie5,Xie1}. The analysis of the     formula for the  distribution type of  synthetic ``collaborators per author" is the same as that in Reference~\cite{Xie6}.  As shown in Fig.~\ref{fig2}c,  the  synthetic  in-degree distribution type is a  mixture of generalized Poisson and power-law, hence the   formula  is analyzed piecewise.
 The formula  for the  head and that for the tail of the type are deduced respectively. The cross-over   can be well fitted  by the formula in the notes of Table~\ref{tab3}.

The  in-degrees   contributed by the second half of  Step 2.b are due to a    random    selection.
  Together with the preset small domain of $f(x)$ in this step, the    effect of the second half  on in-degree distribution is small enough to be ignored, when  compared with that
 contributed by the first half.

The first half of Step 2.b makes the expected in-degree   of a node generated at time $t$ to  be $k^-(t)\approx   {\alpha_l}\delta p T^{ {\beta_l} } t^{- {\beta_l} }/\beta_l-1$, where $l=2,3$ and $\delta=N_1 /2\pi$.
   If $t$ is large enough (suppose larger than a big number $T_1$),  $k^-(t)$ is small enough, and
    changes slowly over $t$.
Hence the formula for the head    is
\begin{align}\label{eq2}P_S(k)&     =  \frac{1}{T-T_1+1} \sum^{T}_{t=T_1}    \frac{k^-(t)^k}{ k!}   \mathrm{e}^{ -  k^-(t) }   ,
\end{align} which is  a mixture Poisson  distribution.  A generalized Poisson distribution can be well fitted by a mixture   Poisson distribution, which can be verified numerically.

 The formula for the tail          is deduced   as follows, where the calculations    are inspired by some of the same general ideas
as explored in the  cosmological networks\cite{Krioukov1}:
 \begin{align}\label{eq3}P_L(k )&=    \frac{1}{T_1 } \int^{T_1 +1}_1 \frac{ k^-( t)^k}{    k! } \mathrm{e}^{-{ k^-( t) }}   d t   \propto  \frac{1}{   k! }   \int^{\delta   \alpha_lp   T^{ {\beta_l} }/\beta_l }_{\delta    \alpha_l (T_1 +1)^{-{\beta_l}}p T^{ {\beta_l} }/\beta_l  }    \tau^{k-1-\frac{1}{\beta_l}} \mathrm{e}^{ -\tau }    d\tau  \notag\\
&\approx  \frac{1}{  k! }   \left(\frac{k-1-\frac{1}{\beta_l}}{\mathrm{e}}\right)^{k-1-\frac{1}{\beta_l}}      \int^{\delta    \alpha_l p  T^{ {\beta_l} }/\beta_l }_{\delta    \alpha_l   (T_1 +1)^{- {\beta_l}}p  T^{ {\beta_l} }/\beta_l }  \mathrm{e}^{\ -\frac{(\tau-k+1+\frac{1}{\beta_l})^2}{2(k-1-\frac{1}{\beta_l})}}   d\tau   \notag\\
&\approx    \frac{\Gamma(k-\frac{1}{\beta_l})}{  \Gamma(k+1) }   \int^{\delta  \alpha_l p  T^{ {\beta_l} }/\beta_l}_{\delta     \alpha_l  (T_1 +1)^{- {\beta_l}} p  T^{ {\beta_l} }/\beta_l } \frac{ \mathrm{e}^{\ -\frac{(\tau-k+1+\frac{1}{\beta_l})^2}{2(k-1-\frac{1}{\beta_l})}} } {\sqrt{2\pi (k-1-\frac{1}{\beta_l})}} d\tau .
\end{align}
 Here    Laplace approximation is used in the third step,  and   Stirling's formula is used in the fourth step.
When $k\gg0$, the integration part in Eq.~\ref{eq3} is  free  of $k$ approximatively, which can be verified as  follows:
\begin{align}\label{eq4}  \frac{d}{dk} \int^{L_2}_{L_1 } \frac{\mathrm{e}^{\ -\frac{(\tau-k+\rho)^2}{2(k-\rho)}}}{\sqrt{2\pi (k-\rho)}}    d\tau  =&\frac{ \mathrm{e}^{-\frac{{\left( L_1  - k+\rho\right)}^2}{2  (k-\rho)}}}{2\, \sqrt{2 \pi (k-\rho)}}\left(1+\frac{L_1  }{k-\rho}\right)   -\frac{ \mathrm{e}^{-\frac{{\left( L_2  - k+\rho\right)}^2}{2  (k-s)}}}{2\, \sqrt{2 \pi (k-\rho)}}\left(1+\frac{L_2 }{k-\rho}\right)
  ,
\end{align}
where  $L_1={\delta  \alpha_l(   T_1 +1)^{- {\beta_l}}} p  T^{ {\beta_l} }/\beta_l$,  $L_2= {\delta \alpha_l   } p T^{ {\beta_l} }/\beta_l$, and $\rho=1+ {1}/{\beta_l}$. This derivative  is approximately equal to  $0$ for    $k\gg0$.
Hence
 \begin{align}\label{eq15}P_L(k )& \propto \frac{\Gamma(k-\frac{1}{\beta_l})}{  \Gamma(k+1) }\approx \frac{1}{ k^{1+\frac{1}{\beta_l}}}\sqrt{\frac{k-\frac{1}{\beta_l}-1}{k}} \left(1-\frac{\frac{1}{\beta_l}+1}{k}\right)^{k-\frac{1}{\beta_l}-1}   \mathrm{e}^{\frac{1}{\beta_l}+1} \approx   \frac{1}{ k^{1+\frac{1}{\beta_l}}}
.
\end{align}
 Stirling's formula is used in the first approximation. The second approximation  holds for $k\gg 0$.
Hence  $P_L(k )$  is approximately  a power-law distribution with   exponent $1+ {1}/{\beta_l}$.
So  we obtain that   the   in-degree distribution tail  of the network generated in Step~2.b is   a mixture of power-law distributions with exponents $1+ {1}/{\beta_1}$ and $1+ {1}/{\beta_2}$ respectively.
Note that in Eq.~\ref{eq2}, the condition $k\gg 0$ does not hold, so the power-law does not emerge in the head of the distribution.

\subsection{Flexibility of the model}

 The provided model has the flexibility of  fitting  empirical data from different  sciences. We have shown that the model can  capture specific features of the empirical  data      PNAS 2007-2015,  the papers of which mainly belong to biological sciences. Here we consider  the  data from physical sciences: the papers of Physical review E published   during  2007-2016 (PRE   2007-2016). The    data   are gathered from   the Web of Science. Authors are identified by   their names on   papers.

   Synthetic   data are  generated through the provided model to capture specific features of   PRE  2007-2016.
   The parameters of the  synthetic data    are listed in  Table~\ref{tab97}. Comparisons on statistic   indicators and distributions are shown in Table~\ref{tab98}  and Fig.~\ref{fig14a} respectively.

  \begin{table*}[!ht] \centering \caption{{\bf The parameters of the synthetic data.} }
\footnotesize\begin{tabular}{l llllllll  } \hline
Network sizes   & $T=5,000$ & $N_1=15$&$N_2=3$& $N_3=1$& $p=0.3$ \\
Influential zones&    $\alpha_1=0.2$& $\alpha_2=0.082$& $\alpha_3=0.02$& $\beta_1=0.43$& $\beta_2=0.44$& $\beta_3=0.44$    \\
$f(k)$ in Step  1.b &   $a=2$& $b=0$& $c=1,855 $&  $d={-3.70}$& $q=0.995$&  $I_1=[2,250]$& $I_2=[11,150]$  \\
$f(k)$ in Step 1.c &  $a=2$& $b=0$& &   & $q=1$&   $I_1=[2,250]$&  \\
$f(k)$ in  Step  2.b &   $a=1.45$& $b=0.45$& $c= 19,424$&  $d={-3.95}$ & $q=0.9999$&  $I_1=[0, 65]$&   $I_2=[20, 65]$   \\
\hline
 \end{tabular}
  \begin{flushleft}
\end{flushleft}
\label{tab97}
\end{table*}

\begin{table*}[!ht] \centering \caption{{\bf Typical statistic   indicators of  the   data.} }
\footnotesize\begin{tabular}{l r r r rr r r r r r r r} \hline
Network&NN&NE    & GCC &AP & MO & PG   &NG & AC  &SC & SC2\\ \hline
  PRE-Citation &24,079&46,061&0.301 & 8.714 &0.892&  0.746 &4,623   &0.272 & 0.317&0.391\\
  Synthetic-Citation &65,805&89,787& 0.053  & 7.225&0.676 &0.795 & 12,169     &0.094 & 0.135 &0.290\\
  PRE-Coauthorship &37,528 &90,711&  0.838 &6.060  &0.950&  0.583 &4,209  &  0.394 \\
  Synthetic-Coauthorship &48,956 &92,870 &0.614 & 6.680 &0.885& 0.837& 2,183 &0.100    \\
\hline
 \end{tabular}
  \begin{flushleft} The meanings of headers  are the same as those in  Table~\ref{tab1}.
\end{flushleft}
\label{tab98}
\end{table*}

%
\begin{figure}
\centering
\includegraphics[height=3.3    in,width=6.2   in,angle=0]{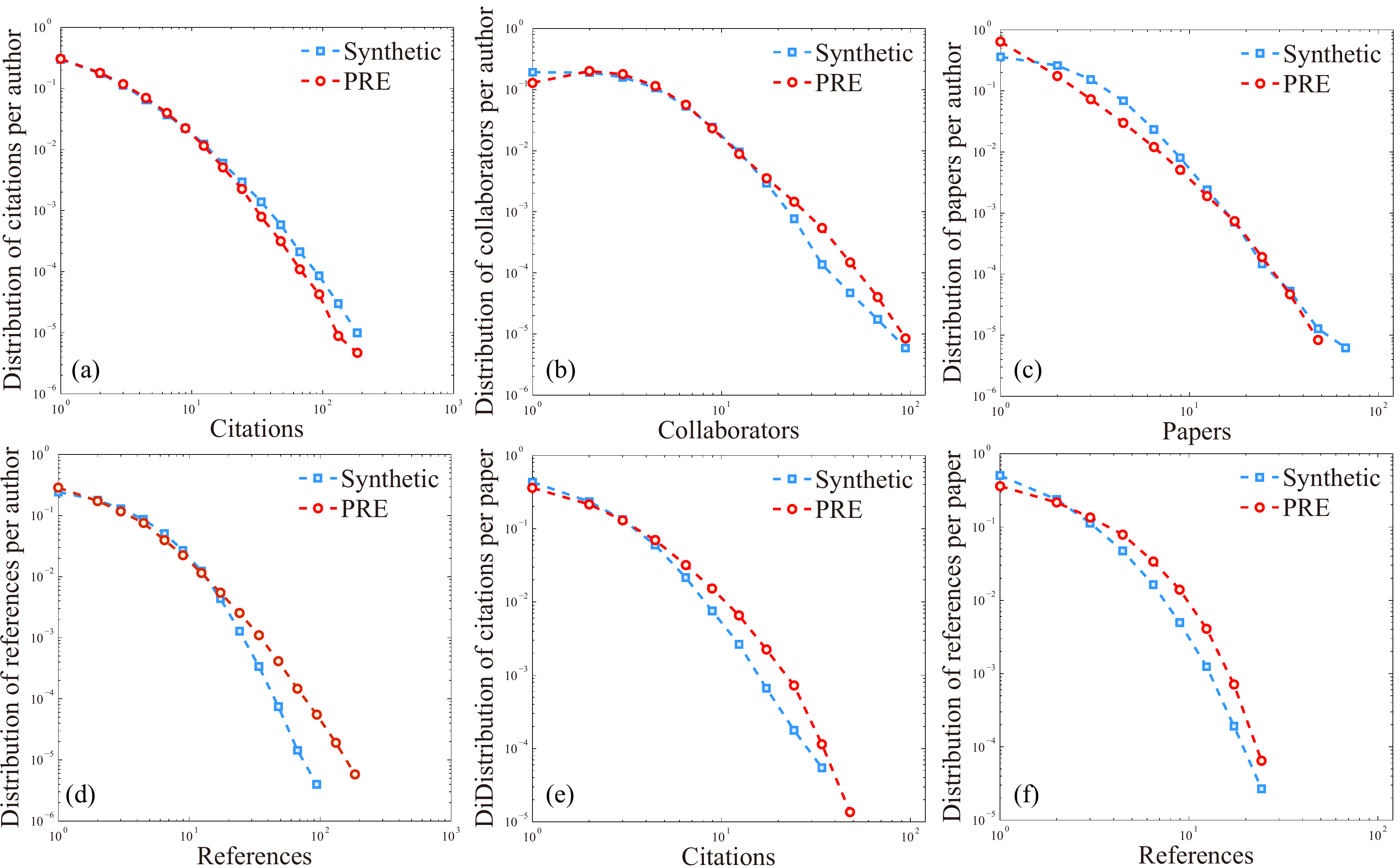}
 \caption{ {\bf The empirical (PRE 2007-2016) and synthetic   distributions of  collaborators/citation/papers/ references per author,  and those of  citations/references per paper.} The data are binned
on abscissa axes  to  show the trends
   hiding in   noise tails. }
 \label{fig13a}      
\end{figure}

\subsection{TARL model}

 Constructively suggested by a reviewer,  the result  of TARL model  is compared with that of the proposed model.
The   pseudo code of  TARL model   in Reference\cite{Borner1}
is repeated as follows.
\begin{description}
\item Initialization
\subitem
Generate $m$ ``papers" and $n$ ``authors" with randomly assigned
``topics";
\subitem  Randomly assign $l$ ``authors" to  the ``papers" within the same ``topic".
\item For time $t=1,2,...,T$ do:
\subitem
  Add $s$ new ``authors" with randomly assigned
``topics";
\subitem Deactivate the ``authors" older than $h$;
\subitem  For each ``topic" do:
\subsubitem Randomly partition the ``authors" within the ``topic" into groups with size $l$;
\subsubitem  For each group  do:
\subsubitem~~~~Randomly read $g$  ``papers" from existing ``papers" within the   ``topic";
\subsubitem~~~~``Select a time-slice form (1 to $t-$1) with probability given in aging-function"\cite{Borner1};
\subsubitem~~~~Generate a new ``paper"  and  randomly  cite $k$ papers (published or cited in this time-slice)  from the read ``papers" and  their references up to  $w$-th level.
\end{description}

\begin{figure}
\includegraphics[height=1.5  in,width=6  in,angle=0]{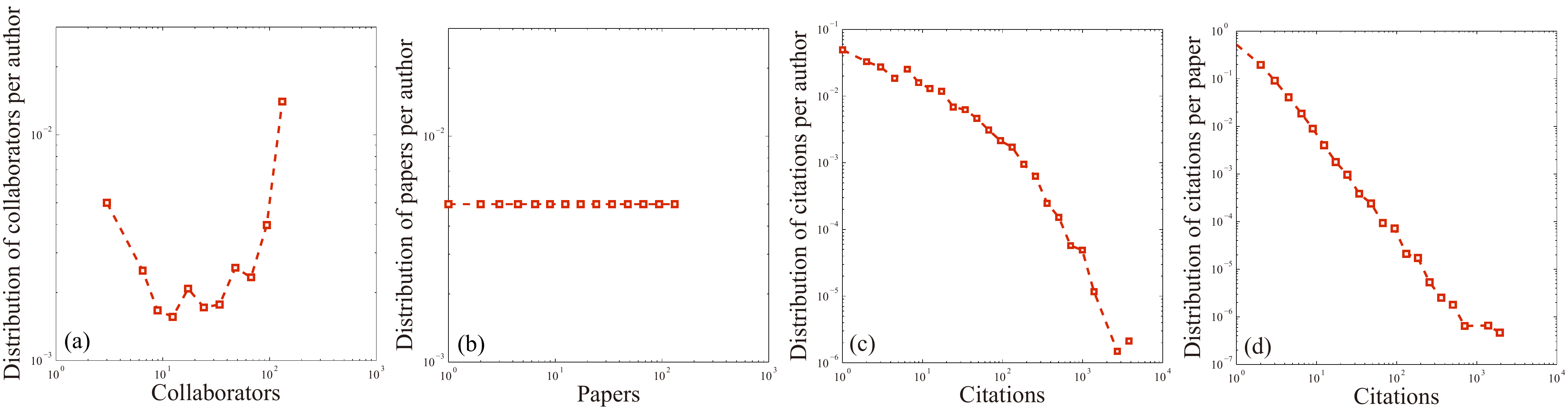}
\caption{  {\bf   Synthetic  distribution  of  collaborators/citations/papers   per author,   and that  of citations   per paper.  } Those distributions are generated through  TARL model.
   } \label{fig14a}   
\end{figure}

The generated
 connections   are restricted to the   ``papers" and ``authors"
within  the same  ``topic".    If no aging-function is
given, then all ``papers" can be  ``read" equally.
  We set the number of ``topics" to be $4$, and no aging-function (so no time-slice). We  let $T=200$, $g=1$, $h=T$, $k=2$, $w=2$, $m=n=l=s=4$.
The generated distribution   of ``collaborators"/``papers"
per ``author"  and that of ``citations" per ``paper"/per ``author" are shown in Fig.~\ref{fig14a}.

 TARL model can generate
a ``coauthorship" network   and  a ``citation" network, which grow simultaneously.
 The ``citation" network is scale-free   (caused by  recursive linking), and has  a positive clustering coefficient    (caused by citing the   ``papers" within the same  ``topic").
Our model harmoniously express  the citation  factors considered in TARL model (i.~e. topics, aging  and recursive follow-up of   citation
references)  by the connection mechanism induced through  the influential zones of   ``papers".
The aging of papers is    expressed by decreasing the sizes of influential zones over $t$.
 In TARL model, ``papers" and ``authors" are assigned specific ``topics" directly.  In our model, we use a continuous way:  expressing nodes' ``topic"   by nodes'  spacial coordinate.
So the circles could be regarded as  ``topic spaces". Note that it is not  a real  topic space, which is a high  dimensional space  representing  textual contents of papers. In our model,
  ``papers" can incompletely ``copy" the references of the ``papers" it cited, which is  induced through the   overlapping of influential zones.



 TARL model neither consider  the Matthew effect  on the number of authors' collaborators nor that on    papers.
In addition, the above instantiation
  assumes  that the number of ``papers" per
``author" is a constant.     Hence, the generated distribution of ``papers" per ``author" and that of  ``collaborators" per ``author"~(Fig.~\ref{fig14a}) have no power-law tails, which emerge in the corresponding distributions from real data~(Fig.~\ref{fig2},
   Fig.~\ref{fig13a}).
    Our model expresses     those  Matthew effects geometrically:    older leaders having a larger influential  zone to   obtain  more ``collaborators" and ``papers".  Therefore, our model can   reproduce those power-law tails.

%







\end{document}